\documentclass[acmsmall,screen,dvipsnames,svgnames]{acmart}
\usepackage[utf8]{inputenc}

\setcopyright{none}
\copyrightyear{2025}
\acmYear{2025}
\acmDOI{}
\usepackage{fancyhdr}
\title{Fair intersection of seekable iterators}

\author{Michael Arntzenius}
\email{daekharel@gmail.com}
\orcid{0009-0002-2234-2549}
\affiliation{%
	\institution{UC Berkeley}
	\city{Berkeley}
	\state{CA}
	\country{USA}
}

\usepackage[frozencache,cachedir=.]{minted}
\setminted{
  xleftmargin=1.5em,
  listparameters=\setlength{\topsep}{0.5em},
}
\usepackage[nameinlink,noabbrev]{cleveref}

\newcommand\hask[1]{\mintinline{haskell}{#1}}
\newcommand\ttt\texttt

\newcommand\todo[1]{{\color{Orange}#1}}
\renewcommand\todo[1]{{\color{IndianRed}#1}}

\newcommand\oldtodo[1]{\todo{#1}}

\renewcommand\todo[1]{\ignorespaces}

\begin{document}

\settopmatter{printacmref=false}
\settopmatter{printfolios=true}
\renewcommand\footnotetextcopyrightpermission[1]{}
\pagestyle{fancy}
\fancyfoot{}
\fancyfoot[R]{\footnotesize\sffamily miniKanren'25}
\fancypagestyle{firstfancy}{
  \fancyhead{}
  \fancyhead[R]{\small\sffamily miniKanren'25}
  \fancyfoot{}
}
\makeatletter
\let\@authorsaddresses\@empty
\makeatother

\begin{abstract}
  miniKanren's key semantic advance over Prolog is to implement a complete yet efficient search strategy, fairly interleaving execution between disjuncts.
  This fairness is accomplished by bounding how much work is done exploring one disjunct before switching to the next.
  We show that the same idea---fairness via bounded work---underlies an elegant compositional approach to implementing worst-case optimal joins using a seekable iterator interface, suitable for shallow embedding in functional languages.
\end{abstract}

\maketitle

\thispagestyle{firstfancy}


\section{Our approach and its scope}
%\section{Scope, approach, audience}
%\section{Background}
%% \section{Who is this paper for?}

%% Relational programming involves both disjunction (union) and conjunction (relational join), but logic programming tends to emphasize the \emph{disjunctive} side (e.g.\ backtracking search), while databases emphasize the \emph{conjunctive} side (e.g.\ indexing, join algorithms).
%% We believe these lines can profitably be blurred.
%% %These lines can and, we believe, should be blurred.
%% %
%% %This paper loosely adapts an idea (\textmu{}Kanren's fair search strategy) from disjunction to conjunction---from logic to database.

This paper originates in early work towards the implemention of a relational programming language inspired by Datalog and modern database query engine techniques, in particular, worst-case optimal joins~\citep{DBLP:conf/pods/NgoPRR12} and specifically Leapfrog Triejoin~\citep{lftj}.
Join algorithms often have set intersection algorithms at their core.
In implementing the `leapfrog' intersection of Leapfrog Triejoin, we discovered it was not compositional---a $k$-way intersection cannot be decomposed into nested binary intersections without loss of performance.

We show how to restore compositionality by borrowing a trick from miniKanren's complete disjunctive search%
%, importing an idea from disjunctive logic programming into conjunctive database querying
.
This permits shallowly embedding efficient multi-way set intersection in a functional language.
Although we do not describe it here, we have extended this to full conjunctive queries.
In future we hope to extend it to full-fledged logic programming.
%In the future we hope to extend this to full conjunctive queries and beyond.
%We believe this technique can be useful to anyone trying to implement database-style relational joins in their favorite functional programming language.
%
We present our work in Haskell because algebraic data types clarify its presentation and laziness is convenient for representing potentially infinite search spaces.
Neither is essential: any disciplined programmer can observe a typing discipline; any language with closures, objects, or similar features can encode laziness.

%% \XXX


\section{Fair union of streams}
%\section{Fair, bounded disjunction}

When multiple rules apply to the current goal, Prolog tries them in order, exploring each to exhaustion before starting the next.
If exploring a rule fails to terminate, later rules are not visited, and potential solutions they might generate are lost.
If we think of goal-directed search as yielding a stream of solutions, Prolog implements disjunction between rules as stream concatenation.
In Haskell, using laziness to represent possibly-infinite streams, we might implement this like so:

\begin{minted}{haskell}
data Stream a = Empty
              | Yield a (Stream a)
append Empty        ys = ys
append (Yield x xs) ys = Yield x (append xs ys)         -- keep focus on xs
\end{minted}

\noindent
As we've hinted, concatenation is an incomplete search strategy: \ttt{append xs ys} will never visit \ttt{ys} if \ttt{xs} is infinite.
We can fix this by interleaving \ttt{xs} and \ttt{ys} instead:

\begin{minted}{haskell}
interleave Empty        ys = ys
interleave (Yield x xs) ys = Yield x (interleave ys xs) -- swap focus to ys
\end{minted}

\noindent
However, this is complete only if both streams are \emph{productive}---they must not loop indefinitely without producing either \hask{Empty} or \hask{Yield}.
%% It's easy to define unproductive streams recursively.
%% For instance, the logic program:
%
%% \begin{minted}{Prolog}
%% theAnswer(X) :- theAnswer(X).
%% theAnswer(42).
%% \end{minted}
%
%% \noindent
%% corresponds, if we implement disjunction via \ttt{interleave}, to an unproductive recursive stream:
%
%% \begin{minted}{haskell}
%% theAnswer :: Stream Int
%% theAnswer = interleave theAnswer (Yield 42 Empty)
%% \end{minted}
%
%% \noindent
Unfortunately, simple and desirable operations on lazy lists may not preserve productivity.
For instance, a productive infinite list becomes unproductive if filtered to a finite subset.
Consider the even prime numbers:

%% filter p Empty = Empty
%% filter p (Yield x xs) = if p x then Yield x xs' else xs'
%%   where xs' = filter p xs
%%
%% primes = Yield 2 (Yield 3 (Yield 5 ... {- definition omitted -}))

\begin{minted}{haskell}
filter :: (a -> Bool) -> Stream a -> Stream a {- definition omitted -}
primes = Yield 2 (Yield 3 (Yield 5 ...)) {- full definition omitted -}
evenPrimes = filter even primes
twoThreeFour = interleave evenPrimes (Yield 3 (Yield 4 Empty))
\end{minted}

\noindent
Here, \ttt{evenPrimes} is equivalent to \mbox{\hask{Yield 2 }$\bot$}, where $\bot$ is an unproductive infinite loop. %TODO
Because of this, \ttt{twoThreeFour} is equivalent to \mbox{\hask{Yield 2 (Yield 3 }$\bot$\ttt{)}}; the call to \ttt{interleave} gets stuck evaluating the $\bot$ in \ttt{evenPrimes} before it can yield \hask{4}.\footnotemark
%This is why \ttt{interleave} is complete only when its arguments are productive.
%In fact, in the presence of $\bot$ no complete union operator is possible for lazy lists \citep[p.\ 85; he calls this ``fair merge'']{clinger-thesis}.

\footnotetext{%
  This incompleteness afflicts the work of \citet{DBLP:conf/icfp/KiselyovSFS05}.
  Their \ttt{interleave} is in effect the CPS translation of ours.
  We implemented \ttt{twoThreeFour} at type \ttt{SFKT Identity Int} and found that \ttt{runIdentity (solve twoThreeFour)} returned \mbox{\ttt{2:3:}$\bot$}.
  More directly, if we let \hask{evenOdds = odds >>- \x -> if even x then return x else mzero}, then \ttt{observe (evenOdds `interleave` return 0)} is $\bot$, but \ttt{observe (return 0 `interleave` evenOdds)} is 0.}

%% TODO: explain semantics, xs = xs ∪ {42}. Usual semantics is xs is the least set satisfying this, so xs = {42}; but *whatever* xs is, we definitely have 42 ∈ xs since 42 ∈ {42} ⇒ 42 ∈ xs ∪ {42} ⇒ 42 ∈ xs. But our approach here won't discover that; so it's not complete.

%% From VolzerVK05:

%% Fairness usually means that a particular choice is taken sufficiently often
%% provided that it is sufficiently often possible [3]. Depending on the
%% interpretation of `choice', `sufficiently often', and `possible', many
%% different fairness notions arise (cf. e.g. [13,4,8]).
%%
%% [3]: "Appraising fairness in languages for distributed programming"
%% [4]: Fairness, Francez, Springer 1986
%% [8]: Kwiatkowska. Survey of fairness notions. Information and Software Technology, 1989. https://wcl.cs.rpi.edu/pilots/library/papers/consensus/survey_Kwiatkowska_1986.pdf
%% [13]: Lehmann, Pnueli, Stavi. Impartiality, justice, and fairnes: etc.
%%
%% Kwiatkowska says: "Intuitively, fairness is a property of computations that can be expressed as follows: no component of the system that becomes possible sufficiently often should be delayed indefinitely."

To find a complete stream union operator, we make an analogy to \emph{fairness} in concurrent programming, which means that ``every process gets a chance to make progress, regardless of what other processes do'' \citep{DBLP:journals/toplas/OwickiL82}%
, although it has no singular formal definition \citep{Kwiatkowska1989,DBLP:conf/concur/VolzerVK05}%
.
%\todo{TODO: SET UP FOR USE OF `BOUNDED FAIRNESS' IN CONJUNCTION. HOW? WHAT'S THE DEFINITION THAT GENERALIZES?}
We can imagine streams as processes that make progress toward finding solutions when evaluated.
In this analogy, stream union is a 2-to-1 multiplexing scheduler that allocates the time given to it between two sub-processes.
Completeness requires fairness: one stream may not permanently block the other from making progress.
Interleaving \emph{elements} between streams is not sufficient, because finding the next element (or determining none exists) can require an infinite amount of work, blocking execution of the other stream.

%In effect, we are doing concurrent programming with cooperative scheduling: processes (streams) must explicitly yield control to other processes (streams).
We therefore need all streams to yield control after doing at most finite work.
Under lazy evaluation, we yield control when we produce a constructor.
%This is why productivity is necessary for fairness.
Therefore, we extend \hask{Stream} with a constructor \hask{Later} that yields control to the consumer without giving further information:

\begin{minted}{haskell}
data Stream a = Empty
              | Yield a (Stream a)
              | Later   (Stream a)
\end{minted}

\noindent
\hask{Later} corresponds to the ``immature'' streams of \textmu{}Kanren~\citep{muKanren}, which it implements as thunks; as Haskell is lazy, we need not explicitly thunk them.
As in \textmu{}Kanren, we actually ensure a stronger property, which we will call \hask{Later}-productivity: all streams do finite work before either terminating or yielding \hask{Later}, disallowing unbroken infinite repetition of \hask{Yield}.
\todo{TODO: intuitive justification for this.}
%Thus \hask{Later} ensures productivity while \hask{Yield} produces new elements.
We can ensure this property conservatively by guarding all recursion with \hask{Later}.\footnotemark{}
For instance, we must redefine \ttt{primes} to be equivalent to \hask{Yield 2 (Later (Yield 3 (Later (Yield 5 ...))))} or similar---%
the exact frequency of \hask{Later} is unimportant so long as it occurs infinitely often.

%% \footnotetext{There is a long line of work on type systems that check productivity by ensuring recursion is appropriately guarded~\citetext{e.g.\ \citealp{DBLP:conf/lics/Nakano00}, \citealp{DBLP:journals/corr/abs-1208-3596},  \citealp{DBLP:conf/icfp/AtkeyM13}}.}

\footnotetext{In many cases it is possible to be less conservative, for instance in our definitions of \ttt{union} and \ttt{filter} below. There is a long line of work on type systems that ensure productivity by checking recursion is appropriately guarded which could be applied here, including \citet{DBLP:conf/lics/Nakano00}, \citet{DBLP:journals/corr/abs-1208-3596}, and \citet{DBLP:conf/icfp/AtkeyM13}.}

Using this, we implement fair stream union (\textmu{}Kanren's \ttt{mplus}):

\begin{minted}{haskell}
union Empty        ys = ys
union (Yield x xs) ys = Yield x (union xs ys)  -- keep focus on xs
union (Later xs)   ys = Later   (union ys xs)  -- swap focus to ys
\end{minted}

\noindent
as well as filter (related to but less expressive than \textmu{}Kanren's \ttt{bind}):

\begin{minted}{haskell}
filter p Empty = Empty
filter p (Yield x xs) = if p x then Yield x xs' else xs'
  where xs' = filter p xs
filter p (Later xs) = Later (filter p xs)
\end{minted}

\noindent
Note that \ttt{union} and \ttt{filter} preserve \hask{Later}-productivity of streams.
Note also that in our concurrency analogy, \ttt{union} acts as a cooperative scheduler, using \hask{Later} as a signal to switch processes.
It's been observed in miniKanren that the placement of immature streams in a program greatly affects performance; this is because it determines the scheduling of (effectively) a massively concurrent program.

Using \ttt{filter} and \ttt{union}, we can define \ttt{twoThreeFour} completely:

\begin{minted}{haskell}
twoThreeFour = union (filter even primes) (Yield 3 (Yield 4 Empty))
\end{minted}

\noindent
If we assume \ttt{primes} is \ttt{Yield 2 (Later ...)}, then \ttt{twoThreeFour} is equivalent to \ttt{\color{Blue} Yield 2 (Later {\color{Red}(Yield 3 (Yield 4 {\color{Blue}$\bot$}))})}, where color indicates source: \ttt{filter even primes} generates the the blue terms \ttt{\color{Blue}Yield 2}, \ttt{\color{Blue}Later}, $\color{Blue}\bot$, while \ttt{Yield 3 (Yield 4 Empty)} generates the red terms \ttt{\color{Red}Yield 3}, \ttt{\color{Red}Yield 4}.

Complete disjunction, then, is accomplished by bounding the work a stream does before handing control to the next stream.
In the rest of this paper, we apply a similar trick to intersection rather than union.
In \cref{sec:unfair-intersection} we show that the leapfrog intersection of seekable iterators used in the worst-case optimal join algorithm Leapfrog Triejoin~\citep{lftj} is \emph{not} fair.
To remedy this, in \cref{sec:fair-intersection} we show how to extend the seekable iterator interface to allow \emph{bounded interleaving} between sub-iterators.
\oldtodo{TODO: rephrase `bounded interleaving'?}
\todo{TODO: need a section where we explain bounded fairness and how that serves as the analogy for the conjunctive fairness we seek to achieve.}






%% \section{Fairness, completeness, productivity}

%% The concept of \emph{fairness} originates in the theory of concurrent processes, meaning roughly that ``every process gets a chance to make progress, regardless of what other processes do'' \citep{DBLP:journals/toplas/OwickiL82}.
%% %
%% Fairness has no one definition; rather, fairness properties form a class \citep{DBLP:conf/concur/VolzerVK05}.

%% \todo{What we really care about is completeness.}

%% \todo{``Interleave'' is fair if its arguments are productive. But it is impossible to ensure productivity of streams while remaining sufficiently expressive: for instance, as soon as we have infinite streams and can filter, we can construct an unproductive stream, \ttt{filter isEven odds}. This is the problem with \citet{DBLP:conf/icfp/KiselyovSFS05}. This is the purpose of \hask{Later}: to allow an iterator to remain productive even if it never generates any elements.}

%% \todo{\emph{Bounded} fairness, not just ``eventually'' but bounding how much \emph{work/time} we do before consulting a stream.}

%% % and \citet{DBLP:journals/jacm/VolzerV12} 




\section{Unfair intersection of seekable iterators}
\label{sec:unfair-intersection}

Suppose we have a collection of key-value pairs, stored in sorted order so that we can efficiently seek forward to find the next pair whose key satisfies some lower bound by e.g.\ galloping search~\citep{DBLP:journals/ipl/BentleyY76}.
We can capture these assumptions in a \emph{seekable iterator} interface:

\begin{minted}{haskell}
data Iter k v = Empty
              | Yield k v (k -> Iter k v)
\end{minted}

\noindent
A seekable iterator, \hask{Iter k v}, is like a stream of key-value pairs, \hask{Stream (k,v)}, except that (a) it yields pairs in ascending key order, and (b) rather than the entire remainder of the stream, \hask{Yield} produces a function \hask{seek :: k -> Iter k v}
%which seeks forward, i.e.\
such that
\ttt{seek target} iterates over the remaining pairs with keys \ttt{k >= target}.
To recover the entire remainder, we pass \ttt{seek} the just-visited key; this lets us iterate over all elements of the stream:

\begin{minted}{haskell}
toSorted :: Iter k v -> [(k, v)]
toSorted Empty = []
toSorted (Yield k v seek) = (k,v) : toSorted (seek k)
\end{minted}

\noindent
We can easily turn a sorted list \hask{[(k, v)]} into a seekable iterator, although seeking will not be efficient since Haskell lists allow only linear, in-order access.
We could use a more appropriate data structure, such as a sorted array or balanced tree, but omit this as it is not crucial to our explanation:

\begin{minted}{haskell}
fromSorted :: Ord k => [(k, v)] -> Iter k v
fromSorted [] = Empty
fromSorted ((k,v) : rest) = Yield k v seek
  where seek k' = fromSorted (dropWhile ((< k') . fst) rest)
\end{minted}

\noindent
We can intersect two seekable iterators by leapfrogging: repeatedly advance the iterator at the smaller key towards the higher, until either both iterators reach the same key or one is exhausted:

\begin{minted}{haskell}
intersect :: Ord k => Iter k a -> Iter k b -> Iter k (a,b)
intersect Empty _ = Empty
intersect _ Empty = Empty
intersect s@(Yield k1 x s') t@(Yield k2 y t') =
  case compare k1 k2 of
    LT -> intersect (s' k2) t -- s < t, so seek s toward t
    GT -> intersect s (t' k1) -- t < s, so seek t toward s
    EQ -> Yield k1 (x,y) (\k' -> intersect (s' k') (t' k'))
\end{minted}

\noindent
%So far, so good.
However, \ttt{intersect}'s performance can suffer asymptotically when intersecting more than two iterators.
For instance, consider three subsets of the integers between 0 and 7,777,777---the evens, the odds, and the endpoints:

\begin{minted}{haskell}
evens = fromSorted [(x, "even") | x <- [0, 2 .. 7_777_777]]
odds  = fromSorted [(x, "odd")  | x <- [1, 3 .. 7_777_777]]
ends  = fromSorted [(x, "end")  | x <- [0,      7_777_777]]
\end{minted}

\noindent
The intersection of \ttt{evens} and \ttt{odds}, and therefore of all three sets, is empty.
We can compute this by calling \ttt{intersect} twice, but performance improves dramatically if, rather than intersecting \ttt{evens} and \ttt{odds} first, we intersect one of them with \ttt{ends} first.
At the GHCi repl:

\begin{minted}{haskell}
ghci> -- set +s to print time statistics
ghci> :set +s
ghci> toSorted ((evens `intersect` odds) `intersect` ends)
[]
(5.57 secs, 5,288,958,128 bytes)
ghci> toSorted (evens `intersect` (odds `intersect` ends))
[]
(0.57 secs, 248,961,040 bytes)
\end{minted}

\noindent
The reason is simple: ``leapfrogging'' \ttt{evens} and \ttt{odds} against one another performs a full linear scan of both lists.
Whatever our current key $x$ in \ttt{even} (e.g.\ $x = 1$), we seek forward past $x$ in \ttt{odds} and reach $x+1$; then we seek past $x+1$ in \ttt{even} to $x + 2$, and so on.
We do 7,777,777 units of work before we determine the intersection is empty.
By contrast, intersecting \ttt{odds} with \ttt{ends} almost immediately skips to the end of both relations.
(This occurs even though we are \emph{not} materializing any intermediate results.)%
\footnote{\label{fn:white-lie}%
  We are telling a white lie here: because we use Haskell lists, seeking is linear-time, and there is no asymptotic slow-down.
  We are instead observing the difference between an \emph{interpreted} inner loop (\ttt{intersect}, loaded at the GHCi repl) and a \emph{compiled} one (\ttt{dropWhile} from the standard library).
  However, had we used sorted arrays with binary or galloping search, or balanced trees, there would be a true asymptotic speed-up for the reasons we describe.
}

The problem is that \ttt{intersect} does not---and with our \hask{Iter} interface, \emph{cannot}---fairly interleave work between its arguments.
Instead, like \ttt{interleave}, it blocks first on one, then the other.
Our leapfrogging \ttt{intersect} implements the $k=2$ case of the $k$-way set intersection of \citet{DBLP:conf/soda/DemaineLM00} and \citet{lftj}.
%% For this reason \citet{DBLP:conf/soda/DemaineLM00} and \citet{lftj} both give $k$-way intersection algorithms for arbitrary $k$.
%, with the expectation that it will be applied ``flat'' to all intersected iterators, rather than ``nested''---the output of leapfrog intersection, despite presenting a seekable iterator interface, cannot be used as input to leapfrog intersection without risking performance loss.
Repeated application of 2-way \ttt{intersect} fails to capture the essence of this $k$-way algorithm, which is \emph{to propagate lower bounds on keys between all intersected iterators.}
%When intersecting two iterators produced by \ttt{fromSorted}, we propagate information back and forth between them: the key we find in the first becomes our target in the second, and vice versa.
Evaluating \ttt{(evens `intersect` odds) `intersect` ends} immediately waits for \ttt{evens `intersect` odds} to find a key, which takes $O(n)$ work (where $n$ is the size of \ttt{evens}/\ttt{odds}).
This blocks information exchange between \ttt{ends} and \ttt{evens}/\ttt{odds} that would let us jump straight to the end in $O(\log n)$ time.

Can we capture this algorithm compositionally, so that repeated 2-way intersection is as efficient as $k$-way intersection? Yes: as with stream union, we can side-step the issue by changing our interface to \emph{bound} how much work we do.

%% However, all is not lost, nor must we move to a $k$-way intersection primitive.
%% As with stream union, we can side-step the issue by changing our interface to \emph{bound} how much work we do.


\section{Fair intersection of seekable iterators}
%\section{Worst-case optimal iteration as bounded, fair conjunction}
\label{sec:fair-intersection}

Just as we implemented fair interleaving of streams by allowing a stream \emph{not} to yield an element, we will implement fair interleaving of seekable iterators by allowing them \emph{not} to yield a key-value pair.
Applying this lesson na\"ively, we might come up with:

\begin{minted}{haskell}
data Iter' k v = Empty
               | Yield k v (k -> Iter' k v)
               | Later     (k -> Iter' k v)
\end{minted}

\noindent
However, this is insufficiently expressive.
The essence of leapfrog intersection is to communicate lower bounds between intersected iterators.
\hask{Iter'} produces lower bounds only by yielding key-value pairs.
Yet if we interrupt a leapfrogging intersection while it is still searching for the next key-value pair, it may still be able to contribute a new lower bound---for instance, while intersecting \ttt{evens} and \ttt{odds}, after \ttt{evens} proposes \ttt{k = 0} and \ttt{odds} seeks forward to reach \ttt{k = 1}, but before we seek again in \ttt{evens}, we already know that \ttt{k >= 1}.
Therefore instead we regard each iterator as having a \hask{Position}, which may be either a key-value pair or a lower bound on future keys:

\begin{minted}{haskell}
data Position k v = Found k v | Bound (Bound k)
data Bound k = Atleast k | Greater k | Done deriving Eq
\end{minted}

\noindent
\hask{Found k v} means we've found a key-value pair \ttt{(k,v)}.
\hask{Bound (Atleast k)} means all future keys are \ttt{>= k}.
\hask{Bound Done} means the iterator is exhausted.
We will see the purpose of \hask{Greater} shortly.

We can now define the type \hask{Seek} of seekable iterators supporting fair intersection, which possess both a position and a seek function:

\begin{minted}{haskell}
data Seek k v = Seek
  { posn :: Position k v          -- a key-value pair, or a bound
  , seek :: Bound k -> Seek k v } -- seeks forward toward a bound
\end{minted}

\noindent
It simplifies the definition of intersection if seeking is idempotent; thus we require \ttt{seek} to consider the remainder of the sequence \emph{including} the current key, rather than dropping it as we did in \hask{Iter}.
%Unlike in \hask{Iter}, we require this seek function to consider the remainder of the sequence \emph{including} the current key, rather than dropping it; thus repeatedly seeking to a given bound is idempotent, which simplifies the definition of intersection.
We must take this into account when defining \ttt{toSorted}:

\begin{minted}{haskell}
toSorted :: Seek k v -> [(k,v)]
toSorted (Seek (Bound Done) _)    = []
toSorted (Seek (Bound p)    seek) = toSorted (seek p)
toSorted (Seek (Found k v)  seek) = (k,v) : toSorted (seek (Greater k))
\end{minted}

\noindent
\oldtodo{TODO: also need to explain the \hask{Bound k} case where we retry!}
When the iterator has \hask{Found} a pair \ttt{(k,v)}, we pass \hask{Greater k} to its seek function to advance \emph{beyond} the key \ttt{k}.
The idea is that \ttt{seek b} seeks towards the first (smallest) key satisfying the bound \ttt{b}.
We already know that \hask{Atleast lo} is satisfied by keys \ttt{k >= lo}.
\hask{Greater lo} is satisfied only by \emph{strictly greater} keys, \ttt{k > lo}.
And \hask{Done} is satisfied by no keys whatsoever.
This endows bounds with a natural order: for bounds \ttt{p,q} we let \ttt{p <= q} iff any key satisfying \ttt{q} must satisfy \ttt{p}.
We can implement this concisely, if nonobviously, as follows:

\begin{minted}{haskell}
satisfies :: Ord k => Bound k -> k -> Bool
satisfies bound k = bound <= Atleast k
instance Ord k => Ord (Bound k) where
  compare x y = embed x `compare` embed y where
    embed (Atleast k) = (1, Just (k, 1))
    embed (Greater k) = (1, Just (k, 2))
    embed Done        = (2, Nothing)
\end{minted}

\noindent
Using \ttt{satisfies} we can define \ttt{fromSorted}:

\begin{minted}{haskell}
fromSorted :: Ord k => [(k,v)] -> Seek k v
fromSorted l = Seek posn seek where
  posn = case l of (k,v):_ -> Found k v
                   []      -> Bound Done
  seek target = fromSorted (dropWhile (not . satisfies target . fst) l)
\end{minted}

\noindent
To define intersection we need one last helper, which extracts a lower bound on the remaining keys in an iterator:

\begin{minted}{haskell}
bound :: Seek k v -> Bound k
bound (Seek (Found k _) _) = Atleast k
bound (Seek (Bound p)   _) = p
\end{minted}

\noindent
Finally, we can define fair intersection of seekable iterators.
If both iterators are at the same key and have found values, we've found an element of the intersection; otherwise we only know there are no keys until after the greater of their bounds.
To seek, we simply seek our sub-iterators.
\oldtodo{Point out that this \emph{bounds} the work done by a single call to seek.}

\begin{minted}{haskell}
intersect :: Ord k => Seek k a -> Seek k b -> Seek k (a,b)
intersect s t = Seek posn' seek' where
  posn' | Found k x <- posn s, Found k' y <- posn t, k == k' = Found k (x, y)
        | otherwise = Bound (bound s `max` bound t)
  seek' k = seek s k `intersect` seek t k
\end{minted}

\noindent
The natural next step is to try our evens-odds-ends example.
Unfortunately, we now run into the ``white lie'' noted in footnote 5 in \cref{fn:white-lie}:
we cannot expect an asymptotic speedup while we are still using linear-access Haskell lists; the actual effects we saw there were due to interpretation overhead.
In a small experiment with Haskell's Arrays package, compiled with \ttt{ghc -O}, on arrays of 30 million elements, our na\"ive \hask{Iter} approach from \cref{sec:unfair-intersection} took 1.2 seconds to enumerate \ttt{(evens `intersect` odds) `intersect` evens}, while using our \hask{Seek} iterators (or reassociating) took less than a millisecond.
You can find the benchmarking code for \hask{Iter} in \cref{fig:iter-benchmark}; the code for \hask{Seek} is nearly the same, except for an altered \ttt{fromSortedArray} method, shown in \cref{fig:seek-fromsortedarray}.

%% %% TODO FIXME: FINISH THIS
%% \todo{TODO: ok now what do I say? Oh, obviously I test this on my previous example: evens, odds, ends.
%%   UNFORTUNATELY, this (a) runs straight into the white lie I told earlier---it's super slow unless we compile it; (b) when we compile, the three-way intersection is indeed faster than (intersect evens odds), which is good, but (c) why should it be? it's the same asymptotic complexity! it appears that dropWhile is ~3.5--4$\times$ faster than ping-ponging between evens \& odds, which is... a larger constant factor than I expected, but okay.
%% }






\begin{figure}
  \begin{minted}[fontsize=\footnotesize]{haskell}
import Data.Array
import Text.Printf
import System.CPUTime
import System.IO (hFlush, stdout)

... -- Definitions of `data Iter`, `toSorted`, and `intersect` from earlier in paper.

n = 30_000_000
evens = fromSortedArray [(x, "even") | x <- [0, 2 .. n]]
odds  = fromSortedArray [(x, "odd")  | x <- [1, 3 .. n]]
ends  = fromSortedArray [(x, "end")  | x <- [0,      n]]

fromSortedArray :: Ord k => [(k,v)] -> Iter k v
fromSortedArray l = go 0 where
  arr = listArray (0, hi) l
  hi = length l
  go lo = if lo >= hi then Empty else Yield k v seek where
    (k, v) = arr ! lo
    seek tgt = go $ gallop ((tgt <=) . fst . (arr !)) (lo + 1) hi

-- galloping search: exponential probing followed by binary search.
-- O(log i) where i is the returned index.
gallop :: (Int -> Bool) -> Int -> Int -> Int
gallop p lo hi | lo >= hi  = hi
               | p lo      = lo
               | otherwise = go lo 1 where
  go lo step | lo + step >= hi = bisect lo hi
             | p (lo + step)   = bisect lo (lo + step)
             | otherwise       = go (lo + step) (step * 2)
  bisect l r | r - l <= 1 = r
             | p mid      = bisect l mid
             | otherwise  = bisect mid r
    where mid = (l+r) `div` 2

printTime :: String -> a -> IO ()
printTime label result = do
  putStr (label ++ ": "); hFlush stdout
  start_ps <- getCPUTime --in picoseconds
  end_ps <- result `seq` getCPUTime
  printf "%0.6fs\n" (fromIntegral (end_ps - start_ps) / (10^12))

{-# NOINLINE time2 #-}
time2 label xs ys = printTime label $ length $ toSorted $ xs `intersect` ys
{-# NOINLINE time3L #-}
time3L label xs ys zs = printTime label $ length $ toSorted $ (xs `intersect` ys) `intersect` zs
{-# NOINLINE time3R #-}
time3R label xs ys zs = printTime label $ length $ toSorted $ xs `intersect` (ys `intersect` zs)

thrice x = do x; x; x
main = do thrice $ time2 "odds & ends" odds ends
          thrice $ time2 "evens & odds" evens odds
          thrice $ time3L "(evens & odds) & ends" evens odds ends
          thrice $ time3R "evens & (odds & ends)" evens odds ends
  \end{minted}
  \caption{Benchmarking intersection for \hask{Iter}}
  \label{fig:iter-benchmark}
\end{figure}

\begin{figure}
  \begin{minted}{haskell}
fromSortedArray :: Ord k => [(k,v)] -> Seek k v
fromSortedArray l = go 0 where
  hi = length l
  arr = listArray (0, hi) l
  go lo = Seek posn seek where
    posn | lo >= hi  = Bound Done
         | otherwise = Found k v where (k,v) = arr ! lo
    seek tgt = go $ gallop (satisfies tgt . fst . (arr !)) lo hi
  \end{minted}
  \caption{fromSortedArray for \hask{Seek}}
  \label{fig:seek-fromsortedarray}
\end{figure}


\section{Related and future work}

Our intersection algorithm descends from the adaptive set intersection of \citet{DBLP:conf/soda/DemaineLM00} by way of Leapfrog Triejoin~\citep{lftj}.
Intersection is a special case of relational join, and the technique described here extends to full conjunctive queries via \emph{nested} intersections.
This is the `trie' in Leapfrog Triejoin: for instance, a finite ternary relation $R \subseteq A \times B \times C$ can be represented as a 3-level trie, i.e.\ a nested seekable iterator \hask{Seek a (Seek b (Seek c ()))}.
By intersecting on each join column successively, we implement a worst-case optimal join.
Worst-case optimal joins therefore seem to be a form of \emph{fair conjunction,} a connection which we are not sure has been noticed previously.

These nested seekable iterators also support disjunction/union, similar to the merge step in mergesort.
We believe it is possible to extend this to existential quantification, essentially a disjunction over an entire trie level, using either a priority queue or a tournament tree~\citep[Chapter 5.4.1]{DBLP:books/lib/Knuth98a} to merge the sub-iterators.
Moreover, by changing the lowest level of the trie from the unit type \ttt{()} to some other type, e.g.\ \hask{Seek a (Seek b (Seek c Int))}, nested seekable iterators support \emph{weighted logic programming}, where tuples of a relation have an associated value.
Conventionally these values are drawn from a semiring: conjunction multiplies, while disjunction and existentials sum.

Together these amount to an implementation of tensor algebra.
Indeed, our \hask{Seek} interface is essentially equivalent to the \emph{indexed streams} which \citet{indexed-streams} use as a tensor algebra compiler intermediate representation.
Indexed streams replace the \hask{Position} of an iterator with three values: its \emph{index} (\emph{key} in our terminology), a \emph{ready flag} indicating whether it has found a value, and its \emph{value} (which should only be accessed when the ready flag is set).
In place of \hask{seek} they have a \emph{skip} function which takes a key and a flag indicating whether to use \hask{Atleast} or \hask{Greater} semantics.
\oldtodo{all their key/index types are bounded so they don't need \hask{Done}---is this accurate?}

To extend to full logic programming (weighted or not), however, we need \emph{recursion,} which is hard to reconcile with the sorted nature of seekable iterators: self-reference means the existence of a larger key might imply the existence of a smaller one, so we may discover a key exists when it's too late to produce it.
One approach might be to introduce an artifical time dimension, e.g.\ to represent a relation by a sequence of seekable iterators, representing a monotonically growing set, with recursion implemented as feedback with delay.
This resembles a na\"ive implementation of Datalog; perhaps semina\"ive evaluation is also possible.


%% ---------- BIBLIOGRAPHY ----------
%\bibliography{on-fairness}

\end{document}